\documentclass[prl,twocolumn,a4paper,aps,superscriptaddress]{revtex4-1}

\usepackage{graphicx}% Include figure files
\usepackage{dcolumn}% Align table columns on decimal point
\usepackage{bm}% bold math
\usepackage{amssymb}
\usepackage{epsfig}
\usepackage{color}

\begin{document}
\title{A Josephson Quantum Electron Pump}
\author{F. Giazotto}
\email{f.giazotto@sns.it}
\affiliation{NEST, Istituto Nanoscienze-CNR and Scuola Normale Superiore, Piazza S. Silvestro 12, I-56127 Pisa, Italy}
\author{P. Spathis}
\affiliation{NEST, Istituto Nanoscienze-CNR and Scuola Normale Superiore, Piazza S. Silvestro 12, I-56127 Pisa, Italy}
\author{S. Roddaro}
\affiliation{NEST, Istituto Nanoscienze-CNR and Scuola Normale Superiore, Piazza S. Silvestro 12, I-56127 Pisa, Italy}
\author{S. Biswas}
\affiliation{NEST, Istituto Nanoscienze-CNR and Scuola Normale Superiore, Piazza S. Silvestro 12, I-56127 Pisa, Italy}
\author{F. Taddei}
\affiliation{NEST, Istituto Nanoscienze-CNR and Scuola Normale Superiore, Piazza S. Silvestro 12, I-56127 Pisa, Italy}
\author{M. Governale}
\affiliation{School of Chemical and Physical Sciences and MacDiarmid Institute for Advanced Materials and Nanotechnology,
Victoria University of Wellington, P.O. Box 600, Wellington 6140, New Zealand}
\author{L. Sorba}
\affiliation{NEST, Istituto Nanoscienze-CNR and Scuola Normale Superiore, Piazza S. Silvestro 12, I-56127 Pisa, Italy}
\maketitle

\textbf{A macroscopic fluid pump works according to the law of Newtonian mechanics and  transfers a large number of molecules per cycle (of the order of $10^{23}$). 
By contrast, a nano-scale charge pump  can be thought as the ultimate miniaturization of a pump, with its operation being subject to quantum mechanics and with only few electrons or even fractions of electrons  transfered per cycle. It generates a direct current in the absence of an applied voltage exploiting the time-dependence of some properties of a nano-scale conductor.
The idea of pumping in nanostructures was discussed theoretically a few decades ago \cite{thouless,pretre,brouwer,zhou}. 
So far, nano-scale pumps have been realised only in system exhibiting strong Coulombic effects \cite{pothier,martinis,pepper1,pepper2,pekola,fuhrer,buitelaar,kaestner}, whereas evidence for pumping in the absence of Coulomb-blockade has been elusive. 
%from the works of Thouless\cite{thouless}, B\"uttiker\cite{pretre} and Brouwer\cite{brouwer},
% experimental evidence for pumping in systems not in the Coulomb-blockade regime has been elusive. 
 A pioneering experiment by Switkes et al. \cite{exp} evidenced the difficulty of modulating in time the properties of an open mesoscopic conductor at cryogenic temperatures without generating undesired bias voltages due to stray capacitances \cite{brouwer01,dicarlo03}.
One possible solution to this problem is to use the ac Josephson effect to induce periodically time-dependent Andreev-reflection amplitudes in a hybrid normal-superconducting system \cite{russo}.
Here we report the experimental detection of charge flow in an unbiased InAs nanowire (NW) embedded in a superconducting quantum interference device (SQUID).
In this system, pumping may occur via the cyclic modulation of the phase of the order parameter of different superconducting electrodes.  
The symmetry of the current with respect to the enclosed magnetic flux \cite{shutenko,moskalets05} and bias SQUID current is a discriminating signature of pumping.
%\change{The symmetry of the current with respect to the enclosed magnetic flux \cite{shutenko,moskalets05} and its dependence on the  bias SQUID current are discriminating signatures of pumping.}
Currents exceeding 20 pA  
are measured at 250 mK, and exhibit symmetries compatible with a pumping mechanism in this setup which realizes a Josephson quantum electron pump (JQEP).}

\begin{figure}[th!]
\includegraphics[width=7.5cm]{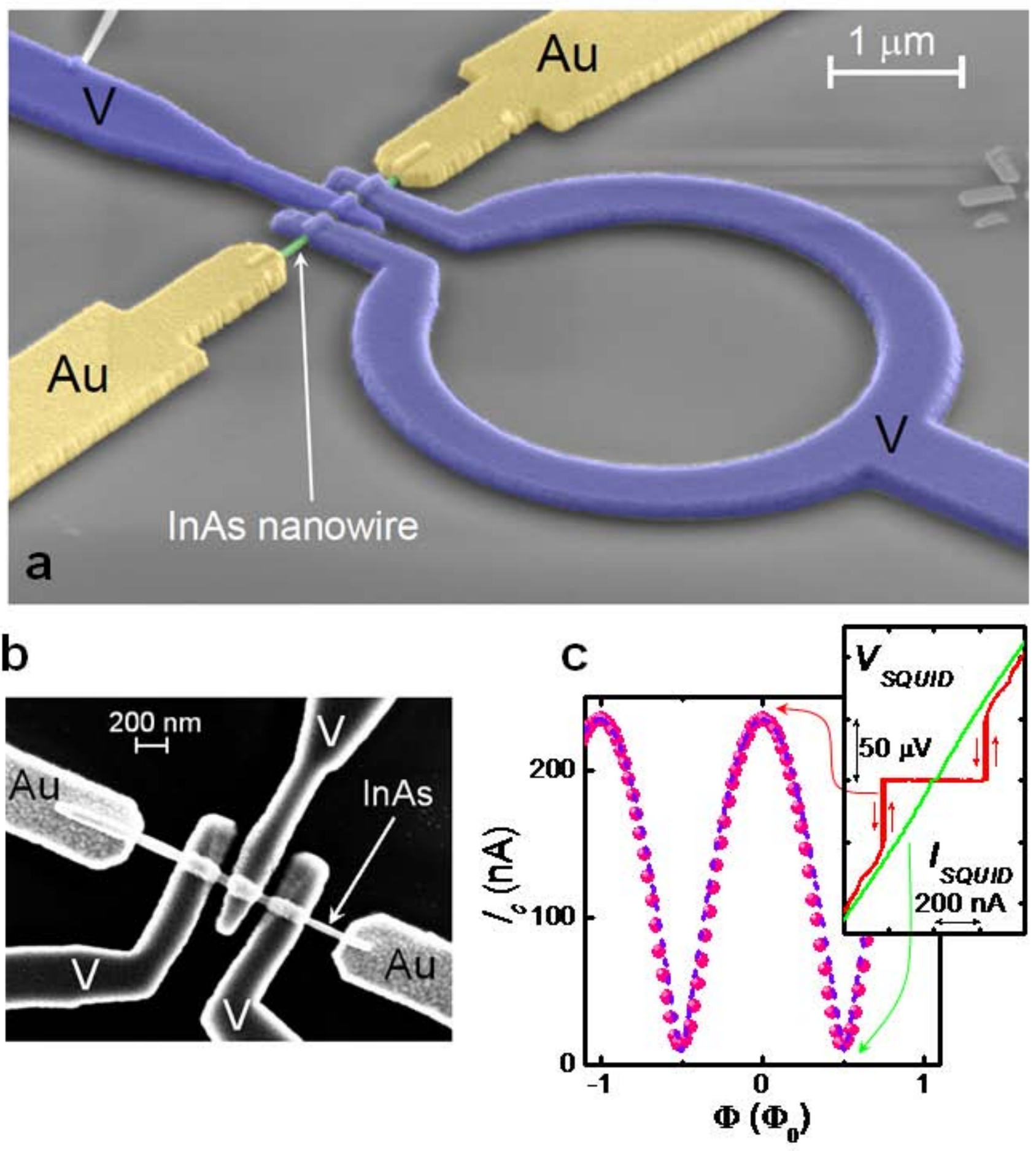}
\caption{\label{fig1} \textbf{InAs Josephson quantum electron pump.} 
(a) Pseudo-color scanning electron micrograph of a JQEP. An InAs nanowire (NW) is connected to three $\simeq 250$-nm-wide V/Ti superconducting contacts forming two $\simeq 50$-nm-long Josephson weak-links and realizing a superconducting quantum interference device (SQUID).
%The internal and external radius of the ring is $\sim 1.75\,\mu$m and $\sim 2.2\,\mu$m, respectively, while
%$\Phi$ is the applied magnetic flux threading the loop.
Two Au/Ti leads, placed at relative distance of $\simeq 1.5\,\mu$m, are contacted to the ends of the NW to allow current detection. The structure was fabricated with electron-beam lithography and evaporation of metals.
The normal-state resistance of the SQUID is $\sim 250\,\Omega$ whereas that of the Au/NW/Au line is $\sim 3.5$ k$\Omega$.
%The distance between the normal metal leads is $\simeq 1.5\,\mu$m.
(b) Blow-up of the device core showing the two V/InAs/V Josephson junctions as well as the two Au electrodes.
(c) Inset:  SQUID voltage ($V_{SQUID}$) versus current ($I_{SQUID}$) characteristics at $\Phi=0$ and $\Phi=\Phi_{0}/2$ ($\Phi$ is the applied magnetic flux whereas $\Phi_{0}$ is the flux quantum) showing a maximum critical current of $\sim 235$ nA. $\Phi_{0}$ corresponds to a magnetic field of $\simeq 1.4$ Oe applied through an effective loop  area of $\sim 14.6\,\mu$m$^2$.
Main panel: $\Phi$-dependent modulation of the SQUID critical current $I_c$. 
Dashed line is the theoretical behavior of a tunnel and resistively-shunted junction SQUID assuming an asymmetry of $\sim 4\%$ between the critical currents of the two  weak-links. Data in (c) are taken at $T=250$ mK.
}
\end{figure}

The microscopic mechanism that enables the transport properties of the NW to be affected by the phases of the superconducting order parameter is Andreev reflection \cite{andreev}.
This is the quantum process for which an electron impinging from the normal side onto the interface between a normal metal and a superconductor,  is retroreflected as a hole (i.e., a time-reversed electron) which picks up the phase of the superconducting order parameter.
%The wave function of the reflected hole and that of the impinging electron have a well defined phase-difference, with the hole picking up the phase of the superconducting order parameter.
When two or more superconductors are connected to the NW, multiple Andreev scattering processes can occur between them so that  transport through the NW will depend on the differences between the phases of the order parameters \cite{SQUIPT}.

The physical realization of this scheme is shown in Fig. 1a 
and consists  of a heavily-doped InAs semiconducting NW on top of which  three fingers of superconducting (S)  vanadium (V) are deposited thus implementing a SQUID \cite{VSQUID}.
Two Au normal-metal electrodes (N) are coupled to the ends of the NW to allow detection of the current $I_{wire}$ flowing through the wire. 
A close-up of the device core is shown in Fig. 1b.
Time-dependence, and possibly pumping, arises from biasing the loop with a current $I_{SQUID}$ larger than the critical current $I_c$ of the SQUID so that the phase differences $\varphi_1(t)$ and $\varphi_2(t)$ across the two Josephson junctions cycle in time at the Josephson frequency $\nu_J=V_{SQUID}/\Phi_0$,
where $V_{SQUID}$ is the voltage developed across the SQUID and $\Phi_0\simeq 2\times 10^{-15}$ Wb is the flux quantum.
In addition, $\varphi_1(t)$ and $\varphi_2(t)$ can be shifted by a constant term $\delta\varphi=2\pi \Phi/\Phi_0$ originating from an applied magnetic flux $\Phi$ threading the loop.
This scheme has the advantage that no high-frequency signal needs to be brought
to the sample thus simplifying the setup and minimizing the impact of stray capacitancies: the time-dependent signal is self-generated thanks to the ac Josephson effect.

Below the critical temperature of the superconductors ($T_c\simeq 4.65$ K) a Josephson current flows through the SQUID across the NW.
The SQUID voltage-current characteristics at 250 mK is shown in the inset of Fig. 1c for two representative values of $\Phi$. Whereas for $\Phi=0$ the characteristic shows a clear dissipationless regime with a critical current $I_c\simeq 235$ nA, for $\Phi=\Phi_0/2$ it behaves almost linearly with $I_c$ largely suppressed. 
The full $I_c(\Phi)$ dependence (main panel of Fig. 1c) shows the characteristic pattern  of a superconducting interferometer. The theoretical curve of a conventional (i.e., described by the RSJ model)  SQUID \cite{tinkham} is shown for a comparison (dashed line, see Supplementary Information).

\begin{figure}[t!]
\includegraphics[width=\columnwidth]{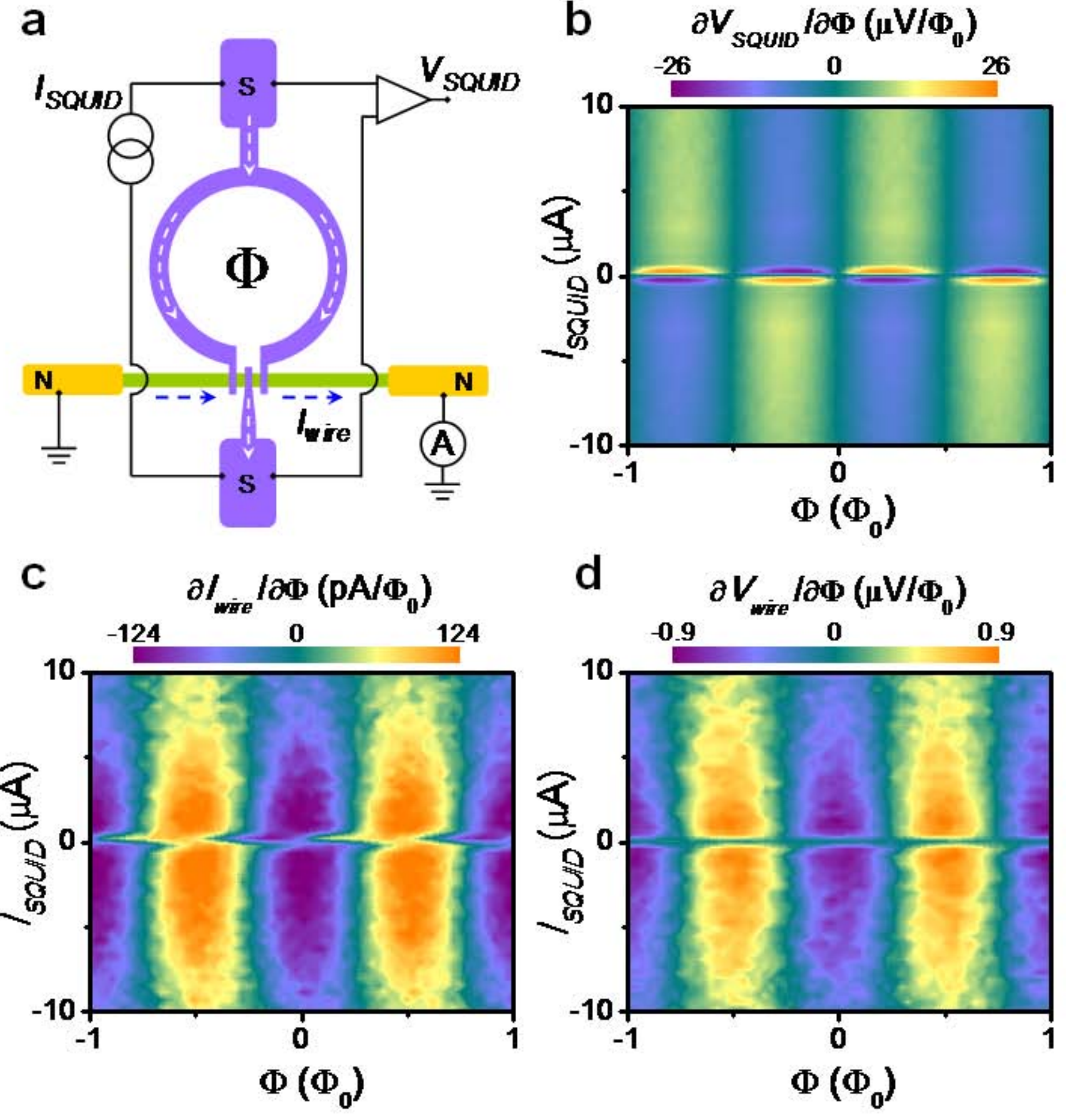}
\caption{\label{fig2} \textbf{Experiment setup and transfer functions characteristics. }
(a) Schematic drawing of the JQEP setup. A dc current $I_{SQUID}$ is fed into the SQUID terminals through a floating source while the  voltage drop $V_{SQUID}$ is recorded against the applied magnetic flux $\Phi$ threading the ring. 
When $I_{SQUID}$ exceeds the SQUID critical supercurrent  the ac Josephson effect sets up inducing a current $I_{wire}$ which flows in the NW. $I_{wire}$ is sensed through an amperometer. 
S and N denote superconductors and normal metals, respectively.
(b) Color plot of the SQUID flux-to-voltage transfer function $\mathcal{V}_{SQUID}=\partial V_{SQUID}/\partial \Phi$ versus $\Phi$ and $I_{SQUID}$.
$\mathcal{V}_{SQUID}$ is antisymmetric in $\Phi$ and $I_{SQUID}$.
(c) Color plot of the NW flux-to-current transfer function $\mathcal{I}_{wire}=\partial I_{wire}/\partial \Phi$ versus $\Phi$ and $I_{SQUID}$.
(d) Color plot of the NW flux-to-voltage transfer function $\mathcal{V}_{wire}=\partial V_{wire}/\partial \Phi$ versus $\Phi$ and $I_{SQUID}$.
Data are taken with a voltmeter in an open-circuit configuration, i.e., without allowing $I_{wire}$ to flow.
Note the markedly different behavior  displayed by $\mathcal{I}_{wire}$ and $\mathcal{V}_{wire}$ which are almost symmetric in $\Phi$ as well as in $I_{SQUID}$.
All measurements are taken at $T=250$ mK with low-frequency phase-sensitive technique to get higher sensitivity and reduced noise. 
}
\end{figure}

Figure 2a shows a sketch of the pumping measurement setup. 
A dc current $I_{SQUID}$ is fed through the SQUID terminals  while the voltage drop $V_{SQUID}$ is measured against $\Phi$. 
The N electrodes are grounded and $I_{wire}$ is sensed with an amperometer.
The N and S parts of the circuit have no common ground therefore preventing any direct net charge transfer
from the SQUID to the NW.

In the following we will concentrate our attention on the symmetries in $\Phi$ and $I_{SQUID}$ displayed by the measured signal, as these are of crucial importance for the interpretation of the experiment.
The low-temperature SQUID flux-to-voltage transfer function $\mathcal{V}_{SQUID}=\partial V_{SQUID}/\partial \Phi$ versus $\Phi$ and $I_{SQUID}$ is displayed in Fig. 2b. 
In particular, $\mathcal{V}_{SQUID}$ is a $\Phi_0$-periodic function of $\Phi$ and is \emph{antisymmetric} in $\Phi$ and $I_{SQUID}$.
By contrast, the flux-to-current transfer function of the NW, $\mathcal{I}_{wire}=\partial I_{wire}/\partial \Phi$ (Fig. 2c), besides exhibiting the same $\Phi_0$-periodicity shows a drastically different behavior, being almost \emph{symmetric} either in $\Phi$ or in $I_{SQUID}$. 
A similar behavior with the same symmetries of $\mathcal{I}_{wire}$ is displayed by the NW flux-to-voltage transfer function, $\mathcal{V}_{wire}=\partial V_{wire}/\partial \Phi$ (Fig. 2d), where $V_{wire}$ is measured with  
open NW contacts.
% i.e. for $\mathcal{I}_{wire}\rightarrow 0$.
$\mathcal{I}_{wire}$ and $\mathcal{V}_{wire}$ result from different but complementary measurements, and the evidence of such a similarity suggests that both reflect the same physical mechanism (see Supplementary Information).
As we shall argue, the nature of the symmetries displayed by $\mathcal{I}_{wire}$ and $\mathcal{V}_{wire}$ is compatible with a quantum pumping mechanisms.

In general, the pumped current is not expected to show definite parity with $\Phi$ \cite{shutenko,moskalets05}, therefore $I_{wire}$ can have a flux-symmetric component  as well. This, however, could be ascribed also to other mechanisms than pumping. In addition, $I_{wire}$ is even not expected to possess any definite parity with $I_{SQUID}$.
In order to extract a pure pumped current contribution from the whole measured signal we focus on the component of $I_{wire}$ which is \textit{antisymmetric} in $\Phi$, $I_{wire}^A$, as it is predicted to be a fingerprint of quantum pumping  in the JQEP \cite{russo}. 
After $\Phi$-integration of $\mathcal{I}_{wire}$, $I_{wire}^A$  is therefore obtained as $I_{wire}^A=[I_{wire}(\Phi,I_{SQUID})-I_{wire}(-\Phi,I_{SQUID})]/2$.
The result of this procedure is shown in Fig. 3a which displays $I_{wire}^A$ versus $\Phi$ and $I_{SQUID}$ at 250 mK. The $\Phi_0$ periodicity joined with the antisymmetry imply that $I_{wire}^A$ vanishes at $\Phi=\Phi_0/2$, while its sign and magnitude can be changed by varying $\Phi$. 
Notably, $I_{wire}^A$ is almost \emph{symmetric} in $I_{SQUID}$. 
The theoretical $I_{wire}^A$ calculated for the JQEP geometry through  a dynamical scattering approach \cite{wang, taddei,blau}  assuming for the NW multiple independent modes is shown in Fig. 3b (see Supplementary Information). 
Although rather idealized, the model is an essential tool to predict the pumped current symmetries of the JQEP. 
Remarkably, summing over many NW modes yields $I_{wire}^A$ which is almost symmetric in $I_{SQUID}$, in agreement with the experiment. 
\begin{figure}[t!]
\includegraphics[width=\columnwidth]{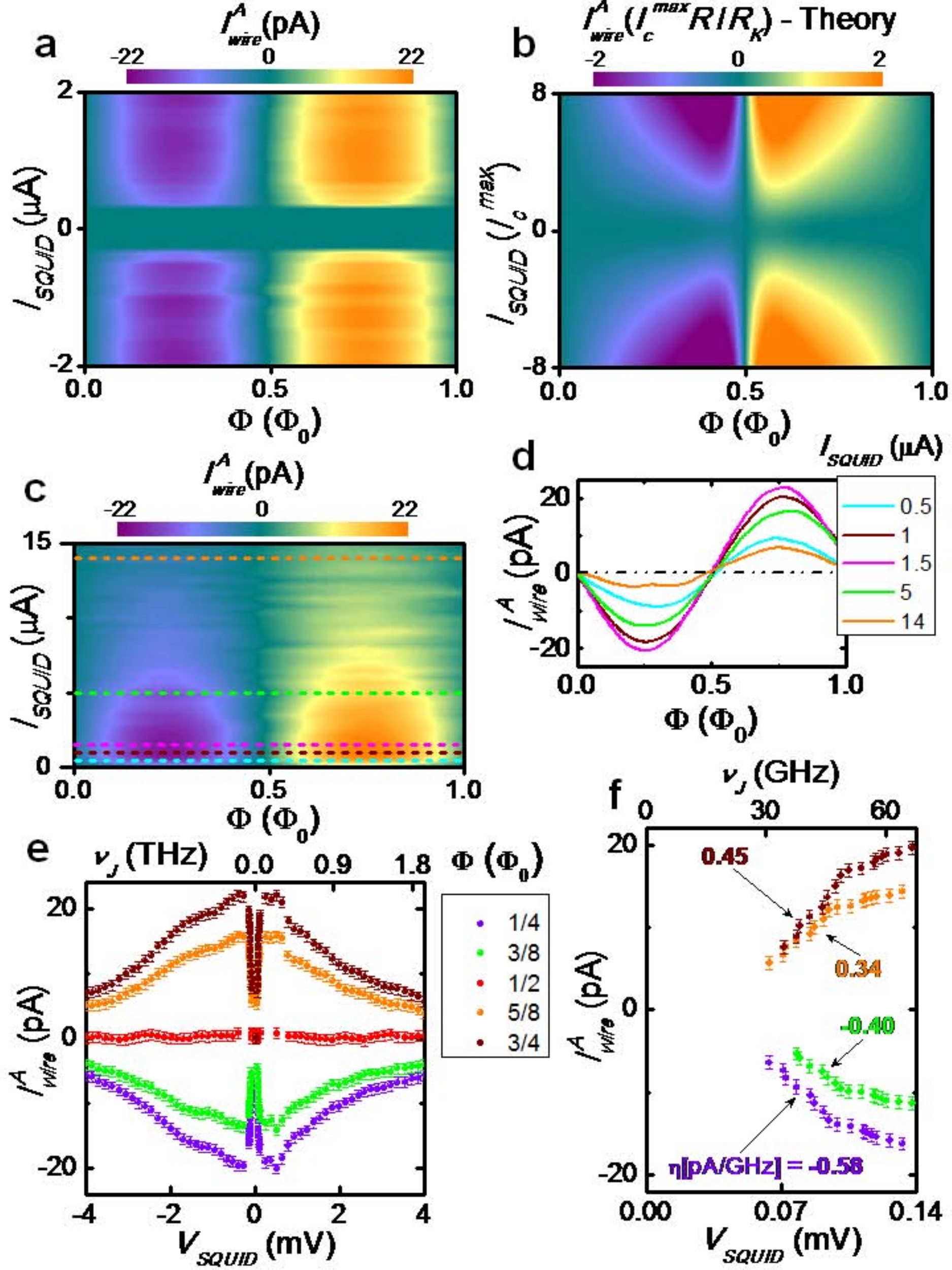}
\caption{\label{fig3} \textbf{Flux and $I_{SQUID}$ dependence of the antisymmetric part of current flowing in the NW. }
(a) Color plot of $I_{wire}^A$ versus $I_{SQUID}$ and $\Phi$.
(b) Color plot of the  theoretical zero-temperature $I_{wire}^A$ versus $I_{SQUID}$ and $\Phi$. 
The calculation was performed for the JQEP geometry assuming the same asymmetry between the Josephson junctions as in the experiment. $I_c^{max}$ is the sum of the critical currents of the two Josephson junctions, $R$ is the total shunting SQUID resistance, and $R_K\simeq 25.8$ k$\Omega$ is the Klitzing resistance (see Supplementary Information for further details).
(c) Color plot of $I_{wire}^A$ versus $I_{SQUID}$ and $\Phi$ shown over a wider range of $I_{SQUID}$. 
(d) $I_{wire}^A$ versus $\Phi$ for a few representative values of $I_{SQUID}$. The latter are indicated as dashed lines of the same color in panel (c).
(e) $I_{wire}^A$ versus $V_{SQUID}$ for a few selected values of $\Phi$.
(f) $I_{wire}^A$ versus $V_{SQUID}$ plotted over a smaller range of $V_{SQUID}$ for the same $\Phi$ values as in panel (e). The slope in the linear regime, expressed in pA/GHz, is denoted with $\eta$. 
In (e) and (f) the error bars represent the standard deviation of the current values calculated over several measurements, and the upper horizontal scale is expressed in terms of the Josephson frequency $\nu_J$. All measurements are taken at $T=250$ mK.
}
\label{fig3}
\end{figure}

Figure 3c shows $I_{wire}^A$ versus $\Phi$ and $I_{SQUID}$ over a wider range of SQUID currents. Specifically, $I_{wire}^A$ turns out to be a non-monotonic function of $I_{SQUID}$, initially increasing then being suppressed for large $I_{SQUID}$. 
This is emphasized in Fig. 3d where $I_{wire}^A(\Phi)$ is plotted for selected values of $I_{SQUID}$. 
$I_{wire}^A$ is a sinusoidal-like function of $\Phi$ whose amplitude depends on $I_{SQUID}$, and is maximized at $\Phi\sim(1/4)\Phi_{0}$ and $\Phi\sim(3/4)\Phi_{0}$. 
\begin{figure}[t!]
\includegraphics[width=\columnwidth]{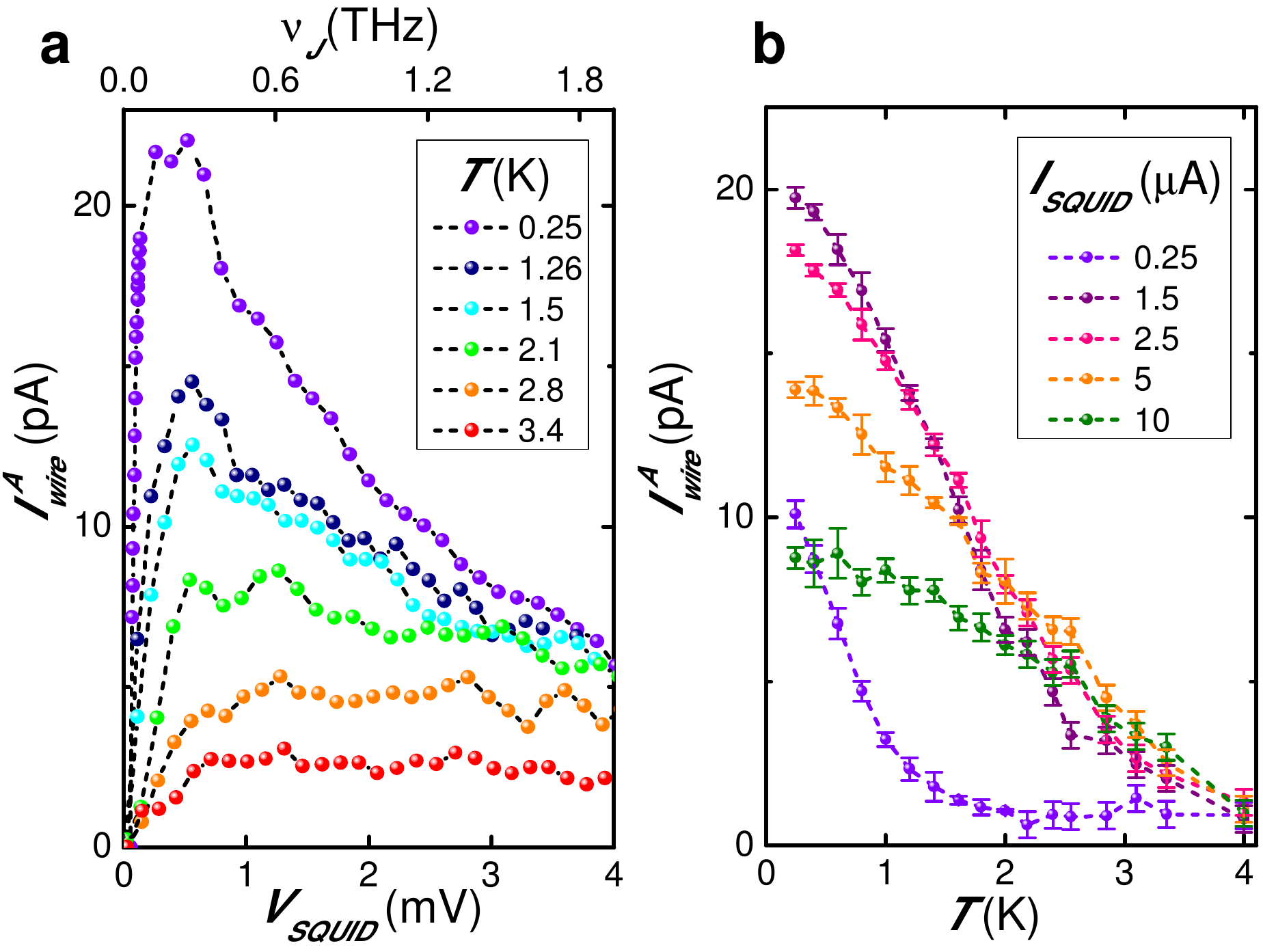}
\caption{\textbf{Temperature dependence of the antisymmetric part of the current flowing in the NW.} 
(a)  $I_{wire}^A$ versus $I_{SQUID}$ measured at several bath temperatures $T$.
(b) $I_{wire}^A$ versus $T$ at selected bias currents $I_{SQUID}$. 
Note the saturation of $I_{wire}^A$ at low temperature as well as its suppression at high $T$. The error bars represent the standard deviation of the current values calculated over several measurements.
Dashed lines in both  panels (a) and (b) are guides to the eye, and all measurements are taken for $\Phi=(3/4) \Phi_0$.
}
\label{fig4}
\end{figure}

The full $I_{wire}^A(V_{SQUID})$ dependence for a few values of flux is displayed in Fig. 3e and highlights both the monotonic 
linear
increase 
for low  $V_{SQUID}$
and suppression at large $V_{SQUID}$. 
The symmetry in $V_{SQUID}$ (i.e., in $I_{SQUID}$) is
emphasized as well.
Furthermore, $|I_{wire}^A|$ is maximized at 
$|V^{max}_{SQUID}|\approx 0.4$ mV independently of $\Phi$, where it
reaches  
values exceeding 20 pA. 
By converting $V_{SQUID}^{max}$ in terms of the Josephson frequency  we get $\nu_J\simeq 190$ GHz whose corresponding time, $\nu_J^{-1}\sim 5$ ps,  is comparable to $\tau_D=W^2/D\simeq 4$ ps, i.e., the time required by electrons to diffuse in the NW between the Josephson junctions. 
In the above expression $W\simeq 250$ nm is the width of the SQUID central electrode (Fig. 1b) which we assume to coincide with
the separation between the weak-links, whereas $D\simeq 0.015$ m$^2/$s is the diffusion coefficient of the NW \cite{JTrans}. 
The transition between the regime of $I_{wire}^A$ enhancement as a function of $V_{SQUID}$ to the one of  $I_{wire}^A$ suppression can be explained  in terms of the ability of the electrons to follow adiabatically the time-dependent parameters up to a maximum frequency set by $\tau_D^{-1}$. Another possible contribution to the suppression observed at larger $V_{SQUID}$ might stem from weakening of the ac Josephson coupling at high applied current \cite{Harris}.

The $I_{wire}^A(V_{SQUID})$ dependence plotted over a reduced bias range is displayed in Fig. 3f. In particular, $I_{wire}^A$ shows a linear behavior with slope $\eta$ which depends on the applied flux, and obtains values as high as several $10^{-1}$ pA/GHz. In the so-called `\emph{adiabatic} regime', i.e., where pumped current is expected to vary linearly with frequency,  $\eta$ would therefore correspond to some $10^{-3}$ electrons per pump cycle.

The role of temperature ($T$) is shown in Fig. 4a which displays $I_{wire}^A$ versus $V_{SQUID}$ at $\Phi=(3/4)\Phi_0$ for several increasing temperatures. $I_{wire}^A$ monotonically decreases upon increasing $T$, which can be ascribed to the influence of thermal smearing as well as thermal-induced dephasing, and is suppressed for  $T\gtrsim 3.5$ K.
We stress that the aforementioned temperature is substantially smaller than $T_c$, the latter setting the disappearance of both Josephson effect and superconductivity in the JQEP.
The $I_{wire}^A(T)$ dependence at the same flux is shown in Fig. 4(b) for a few $I_{SQUID}$ values. Specifically, 
$I_{wire}^A$ begins to round off at lower temperatures indicating a saturation, whereas it is damped at higher $T$. 
Low-temperature behavior suggests that current tends to saturate upon reducing temperature when the ``effective'' separation between 
Josephson junctions 
becomes of the same order of the electron coherence length in the NW, $L_T=\sqrt{\hbar D/(2\pi k_B T)}\sim 270$ nm at 250 mK, where $\hbar$ is the reduced Planck's constant while $k_B$ is the Boltzmann's constant. 
By contrast, the decay of $L_T$ at higher temperatures may be considered as one of the predominant decoherence mechanisms leading to $I_{wire}^A$ suppression. Further study is needed to clarify this point.

It is worthwhile to emphasize that other effects which might manifest in the JQEP would yield currents characterized by symmetries markedly different from the ones predicted for quantum pumping (see Supplementary Information).
Among these we recall (1) any spurious current due to asymmetry between the junctions  %(including the current due to the rectification of an oscillating spurious voltage) 
which is always dominated by a component symmetric in $\Phi$ and antisymmetric in $I_{SQUID}$; (2) any thermocurrent generated by a different power dissipated in the two junctions, which is expected to be predominantly symmetric in both $\Phi$ and $I_{SQUID}$.

We finally note that other normal conductors than InAs NWs could be used for the implementation of the JQEP. This might  pave the way to the investigation of the interplay between superconductivity-induced quantum pumping and exotic electronic states existing, for instance, in graphene \cite{graphene} or in carbon nanotubes \cite{CN}.

We gratefully acknowledge L. Faoro, R. Fazio, L. B. Ioffe, J. K$\ddot{\text{o}}$nig, J. P. Pekola, V. Piazza, H. Pothier, and S. Russo for fruitful discussions, and D. Ercolani for providing the InAs nanowires. The work was partially supported by the NanoSciERA project ``NanoFridge''. F.T acknowledges financial support from EU through the projects ``SOLID'' and ``GEOMDISS''.

\section{Methods summary}

Selenium doped InAs NWs were grown by chemical beam epitaxy on an InAs 111B substrate. Gold catalyst particles were formed by thermal dewetting (at $520^{\circ}$ C for 20 min) of a 0.5-nm-thick Au film under TBA flux. NWs were grown for two hours at $420^{\circ}$ C using TBA, TMI and DTBSe metallorganic precursors with line pressures of 2.0 Torr, 0.3 Torr, and 0.4 Torr, respectively. NWs have diameters of $90\pm 10$ nm and are around $2.5\,\mu$m long. 
Transport parameters were estimated over an ensemble of nominally identical $1\,\mu$m-long NW field effect transistors using a charge control model \cite{Jiang} and a numerical evaluation of the gate capacitance. Carrier density was estimated to be $n=1.8\pm 0.8\times 10^{19}$ cm$^{-3}$ and electron mobility $\mu = 300\pm 100$cm$^2$/Vs. 
The devices were fabricated using a technique of dry cleavage of the NWs onto Si/SiO$_2$ substrates (500 nm oxide on intrinsic Si).
Contacts were obtained by a two-step aligned process: thermal evaporation of Ti/Au (10/80 nm) was performed first and followed by electron-beam deposition of Ti/V (15/120 nm) in an UHV chamber \cite{VSQUID}.  
InAs NWs were treated with a NH$_4$S$_x$ solution before each evaporation step to get transparent metal-NW contacts \cite{JTrans}.

The magneto-electric characterization of the devices was performed in a filtered 
 $^3$He refrigerator (two-stage RC- and $\pi$-filters) down to $\sim 250$ mK using a standard 4-wire technique. 
Current injection at the SQUID terminals was obtained by using a battery-powered floating source, whereas
voltage and current were measured by room-temperature preamplifiers. Derivative measurements (flux-to-voltage as well as flux-to-current transfer functions) were performed with standard low-frequency lock-in technique by superimposing a small modulation to the applied magnetic field. 

\section{Supplementary information}

\subparagraph{Theoretical model}

\begin{figure}[t!]
\includegraphics[width=7cm]{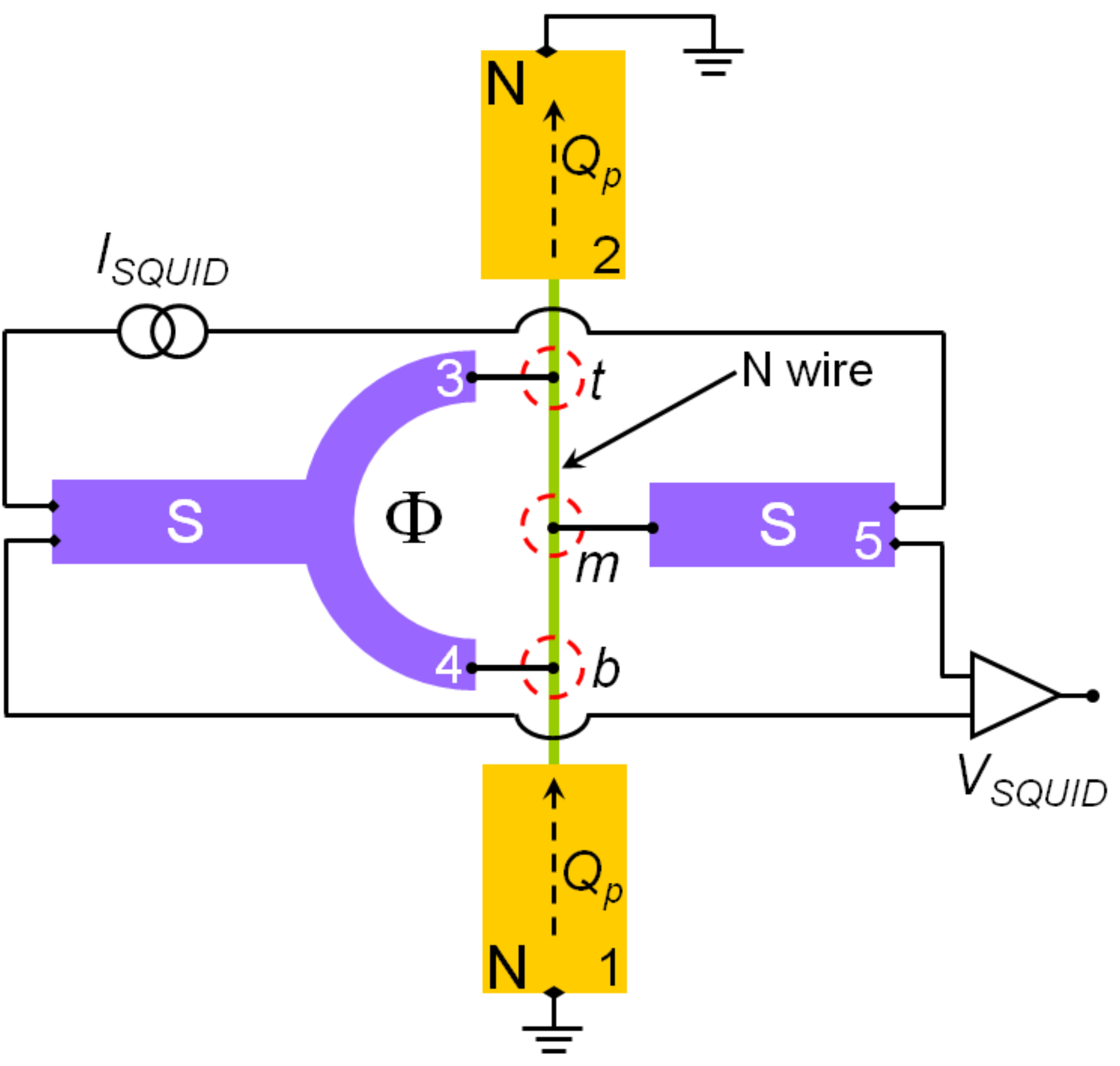}
\caption{\label{fig5} \textbf{Sketch of the Josephson quantum electron pump.}
The system is modeled as three, three-legged beam splitters (denoted by dashed circles) labelled $t$, $m$ and $b$, and connected by two  ballistic normal-metal (N) wires. % of length $L$.
Electrodes 1 and 2 are normal-metallic and grounded, while electrodes 3, 4 and 5 are superconducting (S) and arranged to form a SQUID thread by a magnetic flux $\Phi$. A dc current $I_{SQUID}$ is fed into the SQUID terminals through a floating source determining a voltage drop $V_{SQUID}$. $Q_{p}$ represents the charge pumped per cycle in the N electrodes.
} 
\end{figure}

Let us consider the system depicted in Fig.~\ref{fig5}, where a NW is connected to two normal-metal (N) leads, labelled by 1 and 2, and to three superconducting (S) leads, which implement a SQUID, labelled by 3, 4, and 5.
If the SQUID is polarized by a current $I_{SQUID}$ larger than its critical current, the ac Josephson effect sets in introducing a time-dependence in the scattering amplitudes through the NW and enabling pumping to occur. A magnetic flux $\Phi$ threads the SQUID introducing a phase shift $\delta\varphi=2 \pi \Phi/\Phi_0$, where $\Phi_0=\pi\hbar/e$ is the flux quantum, $e$ is the electron charge and $\hbar$ the reduced Planck's constant.
In the adiabatic regime, the charge pumped per cycle in one of the N leads $Q_i$ (with $i=1,2$) can be calculated in the scattering approach through a generalization of the Brouwer's formula~\cite{brouwer} to hybrid systems~\cite{wang,blau,taddei}.
If $x_1(t)$ and $x_2(t)$ are the two pumping parameters varying along a closed path in the $(x_1,x_2)$-space, at zero temperature one finds that
\begin{equation}
\label{qpump}
Q_i=\frac{e}{\pi}\int_\Omega dx_1 dx_2\sum_{j=1}^2 \Pi_{i,j}(x_1,x_2),
\end{equation}
where $\Omega$ is the area enclosed by the path in parameter space, and
\begin{eqnarray}
\label{Pi}
\Pi_{i,j}(x_1,x_2)=\Im \left\{
\frac{\partial[S_{\text{ee}}^\star]_{i,j}}{\partial x_1}
\frac{\partial[S_{\text{ee}}]_{i,j}}{\partial x_2}-\right. \nonumber\\
\left. -\frac{\partial[S_{\text{he}}^\star]_{i,j}}{\partial x_1}
\frac{\partial[S_{\text{he}}]_{i,j}}{\partial x_2}
\right\} .
\end{eqnarray}
In Eq.~(\ref{Pi}), $S_{\text{ee}}$ and $S_{\text{he}}$ are, respectively, the normal and the Andreev scattering matrices between the two N leads evaluated at the Fermi energy.
Assuming that all leads support a single propagating channel, $[S_{\text{ee}}]_{i,i}$ ($[S_{\text{he}}]_{i,i}$) is the amplitude for an electron entering from lead $i=1,2$ to be reflected back as an electron (hole), while $[S_{\text{ee}}]_{j,i}$ ($[S_{\text{he}}]_{j,i}$), with $j\neq i$, is the transmission amplitude for an electron entering from lead $i$ and exiting through lead $j$ as an electron (hole).
$S_{\text{ee}}$ and $S_{\text{he}}$ can be determined by the scheme proposed in Ref.~\cite{taddei}, which requires the calculation of the scattering matrix $S$ of the system depicted in Fig.~\ref{fig5} when all contacts are in the normal state.
$S$, in turns, is computed as a composition of three (three-legged) beam splitters, indicated by dashed circles in Fig.~\ref{fig5} and labelled by the index $\lambda=\{t,m,b\}$, connected to each other through a pair of ballistic N wires of different lengths.
The scattering matrix of beam splitter $\lambda$ can be written as 
\begin{equation}
S_\lambda=\left(
\begin{array}{ccc}
-\sqrt{1-2\gamma_\lambda}\,e^{i \psi_\lambda} & \sqrt{\gamma_\lambda} & \sqrt{\gamma_\lambda} \\
\sqrt{\gamma_\lambda} & \sqrt{\frac{1-\gamma_\lambda}{2}}\, e^{i\alpha_\lambda} & \sqrt{\frac{1-\gamma_\lambda}{2}} \, e^{i\beta_\lambda} \\
\sqrt{\gamma_\lambda} & \sqrt{\frac{1-\gamma_\lambda}{2}} \, e^{i\beta_\lambda} & \sqrt{\frac{1-\gamma_\lambda}{2}}\, e^{i\alpha_\lambda} 
\end{array}
\right),
\label{esse}
\end{equation}
where $\gamma_\lambda$  takes values between 0 and $1/2$, $\alpha_\lambda=-\psi_\lambda+q_\lambda \arccos [-\gamma_\lambda/(1-\gamma_\lambda)]$ and $\beta_\lambda=-\psi_\lambda-q_\lambda\arccos [-\gamma_\lambda/(1-\gamma_\lambda)]$ with $q_\lambda=\pm1$.
The three S leads, described by constant pair potentials $\Delta_i=|\Delta|\text{exp}(i\phi_i)$ (with $i=3,5$), are assumed to be ideally coupled to the structure so that perfect Andreev reflection occurs at the S interfaces.
When the bias current $I_{SQUID}$ is larger than the critical current of the SQUID, a voltage $V_{SQUID}$ develops across the latter. 
For the SQUID we assume the RSJ  voltage-current relation~\cite{tinkham}
\begin{equation}
V_{SQUID}(\delta\varphi)=\text{sign}(I_{SQUID})R\sqrt{I_{SQUID}^2-I_c(\delta\varphi)^2} ,
\end{equation}
where $R$ is the total shunting SQUID resistance and $I_c(\delta\varphi)$ is the flux-dependent SQUID critical current.
The latter, used to fit the data in Fig. 1c, can be written as
\begin{equation}
I_c(\delta\varphi)=(I_{c1}+I_{c2})\sqrt{r^2+(1-r^2)\cos^2 (\delta\varphi/2)} ,
\end{equation}
where $I_{c1}$ and $I_{c2}$ are the critical currents of the individual Josephson junctions composing the SQUID, and $r=(I_{c1}-I_{c2})/(I_{c1}+I_{c2})$ is the degree of  asymmetry of the SQUID.

From a practical point of view, we first calculate $Q_1$ and $Q_2$ through Eq.~(\ref{qpump}) assuming that the N leads 1 and 2 and lead 5 are grounded, while S leads 3 and 4 are kept at the potential $V_{SQUID}$. This choice sets the phases of the superconductors as follows: 
\begin{eqnarray}
\phi_3&=&\text{sign}(I_{SQUID})\omega_J t \\
\phi_4&=&\text{sign}(I_{SQUID})\omega_J t-\delta\varphi \\
\nonumber
\phi_5&=&-\frac{\delta\varphi}{2}-\arctan\left[ \frac{1}{r}\cot\left(\frac{\delta\varphi}{2}\right)\right]\\
& &+\frac{\pi}{2}\text{sign}(\sin\frac{\delta\varphi}{2}) 
-\frac{\pi}{2} \,\text{sign}(I_{SQUID}), 
%\,\text{sign}(\cos\frac{\delta\varphi}{2}) 
\end{eqnarray}
where $\omega_J=2\pi \nu_J=2\pi |V_{SQUID}| / \Phi_0$ is the Josephson angular frequency and the function $\arctan$ takes values between $-\pi/2$ and $\pi/2$.
The value of $\phi_5$ is chosen to ensure that the supercurrent is maximized in the limit of $\omega_J\to 0$. 
As a consequence of this, all observable quantities exhibit the standard $\Phi_0$ periodicity.
The pumping parameters are defined as $x_1(t)=\cos (\omega_J t)$ and $x_2(t)=\sin (\omega_J t)$ so that $\text{exp}(i\phi_3)=x_1+ix_2$, $\text{exp}(i\phi_4)=(x_1+ix_2)\text{exp}(-i\delta\varphi)$. 
%and $\text{exp}(i\phi_5)=\text{exp}[-i(\delta\varphi/2+\pi/2)]$.
From this choice is clear that the two parameters are maximally out of phase, independently of $\delta\varphi$, and that the path is a circle of radius one centered around the origin.
Since the N and S parts of the circuit have no common ground, the actual chemical potentials of the S electrodes with respect to the potential of the N electrodes have to arrange themselves so that no net current flows between the two parts of the circuit.
As a consequence, the charge pumped per cycle can be written as
\begin{equation}
Q_{p}(\delta\varphi)=\frac{Q_1(\delta\varphi) G_2(\delta\varphi)-Q_2(\delta\varphi) G_1(\delta\varphi)}{G_1(\delta\varphi)+G_2(\delta\varphi)} ,
\label{pumpcurrent}
\end{equation}
where $G_i=\left|[S_{\text{he}}]_{i,1}\right|^2+\left|[S_{\text{he}}]_{i,2}\right|^2$ is the conductance (in units of $e^2/\pi\hbar$) relative to lead $i$.
Note that, in general, $Q_{p}(\delta\varphi)$ has no definite parity in $\delta\varphi$, in agreement with the results of Refs.~\cite{shutenko,moskalets05}, and no definite parity in $I_{SQUID}$.
The $\delta\varphi$-antisymmetric component of the pumped current is obtained as $I^A_{wire}=(\omega_J/2\pi)\ [Q_{p}(\delta\varphi)-Q_{p}(-\delta\varphi)]/2$. 
In Fig.~ 3b $I^A_{wire}$ is plotted in units of $I_c^{max} R/R_{K}$, where $I_c^{max}=I_{c1}+I_{c2}$ and $R_K=2\pi\hbar/e^2$ is the Klitzing resistance. The 
current has been computed assuming that the NW carries  50 independent channels, each of which described by a scattering matrix obtained taking  $\psi_{\lambda}$, $q_{\lambda}$ and the phases accumulated along the two N wires as random parameters, while setting $\gamma_t=1/10$, $\gamma_m=1/11$ and $\gamma_b=1/13$.

In the configuration where lead 1 is a voltage probe (rather than connected to ground) one can calculate the voltage $V_{p}$ which develops at lead 1 as a consequence of the charge pumped. $V_{p}$, determined by setting to zero the current flowing in the NW, can be written as
\begin{equation}
V_{p}(\delta\varphi)=|V_{SQUID}(\delta\varphi)| 
\frac{G_1(\delta\varphi) + G_2(\delta\varphi)}{G_1(\delta\varphi) G_2(\delta\varphi)}  Q_{p}(\delta\varphi).
\end{equation}
The $G_i$ has, in general, no definite parity in $\delta\varphi$ and $I_{SQUID}$. 
Furthermore, $\frac{G_1(\delta\varphi) + G_2(\delta\varphi)}{G_1(\delta\varphi) G_2(\delta\varphi)}$ is approximately even in both quantities also in the presence of a small asymmetry between the two Josephson junctions (which is typically the case of any realistic situation), 
so that $V_p$ and $Q_p$ show the same parity both in $\delta\varphi$ and $I_{SQUID}$. The flux-antisymmetric component of $V_{p}$ is defined as
$V^A_{wire}(\delta\varphi)=[V_{p}(\delta\varphi)-V_{p}(-\delta\varphi)]/2$.

We shall further discuss the spurious effects which can occur in the presence of a shunting dissipative current across the Josephson weak-links. 
If the two Josephson junctions are not equal, a spurious voltage $V_s$ (containing a constant and a time-oscillating component) arises in the NW between the beam splitters  $t$ and $b$ in Fig.~\ref{fig5}. 
This produces a current $I_{s}$ in the NW  that is not originated by quantum pumping. 
On the one hand, the current $I_{s,const}$ related to the constant component of $V_s$ reverses by changing the sign of $I_{SQUID}$, in contrast to $I_{wire}^{A}$, and it is an even function of $\delta\varphi$. 
On the other hand, it turns out that the quantum rectified current $I_{s,rect}$ associated to the oscillating component of $V_s$ has no definite parity both in  $\delta\varphi$ and  $I_{SQUID}$, similarly to $Q_p$ of Eq. (\ref{pumpcurrent}). 
However, 
$I_{s,rect}$ exists only in the presence of a finite  $I_{s,const}$, since they have the same physical origin.
Yet, $I_{s,rect}$ is smaller than $I_{s,const}$ because the amplitude of   
the oscillating components of $V_s$ is  set by $I_{c}^{max}$, whereas the constant component of $V_s$ is proportional to $V_{SQUID}$. 
Therefore, the total spurious current is dominated by the component that is even in flux and odd in $I_{SQUID}$ which would be detected, if present, in the transfer function $\mathcal{I}_{wire}$.  
Since the measured derivative signal $\mathcal{I}_{wire}$ is almost flux-symmetric [see Fig. \ref{fig2}(c)], we can rule out the presence of $I_{s,const}$ and therefore of quantum rectification. 
We stress that even in the presence of a sizable $I_{s,const}$, our calculations predict  $I_{wire}^{A}$ to be typically several orders of magnitude larger than the flux-antisymmetric component of $I_{s,rect}$ (which is even in $I_{SQUID}$) thus fully dominating the measured signal.
\begin{figure}[t!]
\includegraphics[width=\columnwidth]{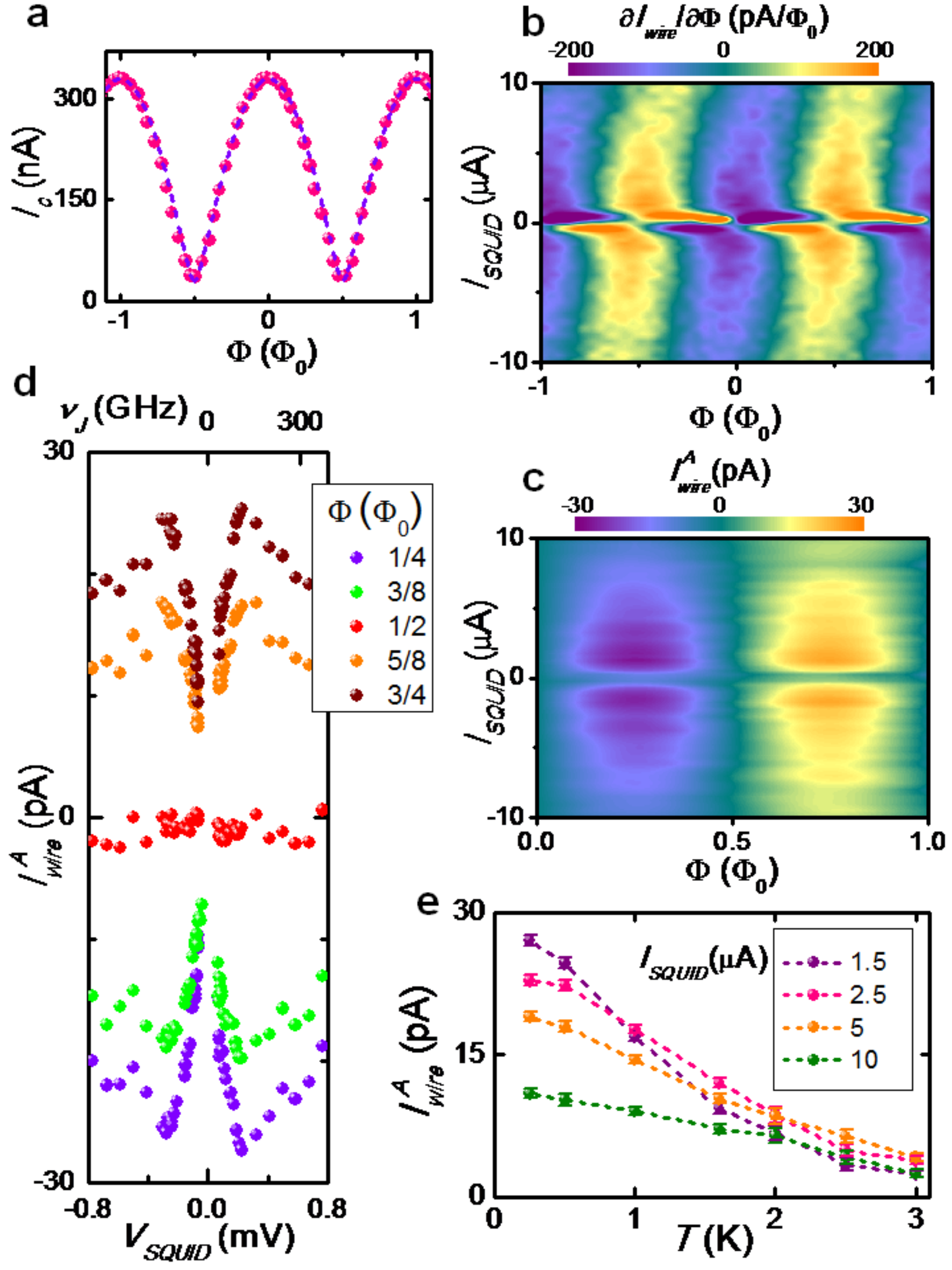}
\caption{\label{fig6} \textbf{Experimental data for a different JQEP.}
(a) $\Phi$-dependent modulation of the SQUID critical current $I_c$. Dashed line is the theoretical behavior of a tunnel and resistively-shunted junction SQUID assuming an asymmetry $r\sim 9\%$ between the critical currents of the two weak-links.
(b) Color plot of the NW flux-to-current transfer function $\mathcal{I}_{wire}=\partial I_{wire}/\partial \Phi$ versus $\Phi$ and $I_{SQUID}$.
(c) Color plot of $I_{wire}^A$ versus $I_{SQUID}$ and $\Phi$.
(d) $I_{wire}^A$ versus $V_{SQUID}$ for a few selected values of $\Phi$. Data in (a)-(d) are taken at $T=250$ mK.
(e) $I_{wire}^A$ versus temperature $T$ at selected bias currents $I_{SQUID}$ for $\Phi=(3/4) \Phi_0$. The error bars represent the standard deviation of the current values calculated over several measurements,
and dashed lines are guides to the eye.} 
\end{figure}

In analogy, the current $I_{SQUID}$ might produce a different power dissipated between points $t$ and $b$ in Fig.~\ref{fig5} leading to a thermocurrent flowing through the NW. 
Since $V_s$ is dominated by its constant component, this thermocurrent would be almost \textit{symmetric} both in $\delta\varphi$ and $I_{SQUID}$, in contrast to $I_{wire}^A$. 
In addition, there could be a small contribution to the thermocurrent due to the oscillating component of $V_s$ which would have no definite parity
both in $I_{SQUID}$ and $\delta\varphi$. Since the power dissipated is proportional to  $V_{s}^2$ such contribution to the thermocurrent is \emph{a fortiori} negligible.

In conclusions, all the mechanisms envisioned above to produce a  spurious  dc current can be distinguished from quantum pumping  by their parity  with respect to magnetic flux  $\Phi$ or  bias current $I_{SQUID}$.

\subparagraph{Supplementary data}

Here we present additional data for another JQEP device with nominally-identical geometry. 
Its essential parameters are the SQUID normal-state resistance of $\sim 187\,\Omega$ and the resistance of the Au/NW/Au line of $\sim 2.1$ k$\Omega$. 
The general behavior of this device is similar to that discussed in the main text although it is characterized by less symmetry between the two Josephson junctions.
Figure \ref{fig6} (a) displays the full $I_c(\Phi)$ dependence of the SQUID measured at 250 mK which shows a maximum critical current of $\sim 330$ nA. Superimposed for a comparison (dashed line) is the model for a tunnel and resistively-shunted junction SQUID \cite{tinkham} assuming an asymmetry $r\sim9\%$ between the critical currents of the two weak-links. 
The low-temperature flux-to-current transfer function $\mathcal{I}_{wire}=\partial I_{wire}/\partial \Phi$ versus $\Phi$ and $I_{SQUID}$ is shown in Fig. \ref{fig6}b. 
$\mathcal{I}_{wire}$ shows no definite parity both in $\Phi$ and $I_{SQUID}$ which stems from the presence of a spurious current $I_{s}$ in the NW, and might be attributed to the reduced symmetry of the SQUID junctions.
Figure \ref{fig6}c shows the extracted $I_{wire}^A$ versus $\Phi$ and $I_{SQUID}$ at 250 mK which highlights both the non-monotonic dependence and symmetry in $I_{SQUID}$.
The full $I_{wire}^A(V_{SQUID})$ dependence for a few selected values of $\Phi$ at 250 mK is displayed in Fig. \ref{fig6}d, and emphasizes the overall symmetry in $V_{SQUID}$. For the present device $|I_{wire}^A|$ is maximized at $|V_{SQUID}^{max}|\approx 0.25$ mV where it obtains values exceeding $\sim 27$ pA. 
$|V_{SQUID}^{max}|$ corresponds to a Josephson frequency $\nu_J\simeq 120$ GHz (and related time $\nu_J^{-1}\sim 8$ ps). This difference from the device presented in the main text could originate from a slightly larger width $W$ of the SQUID central electrode combined with a reduced NW diffusion constant which lead to an increased diffusion time $\tau_D$.
The $I_{wire}^A(T)$ dependence at $\Phi=(3/4)\Phi_0$ is shown in Fig. \ref{fig6}e for a few selected $I_{SQUID}$ currents. 
Specifically, $I_{wire}^A$ is rounded off at low temperature, whereas it is strongly damped and suppressed for $T\gtrsim 3$ K.
The general behavior of $I_{wire}^A$ and the arguments of the previous section  therefore suggest that in this sample $I_{wire}^A$ is fully dominated by quantum pumping, although a small component of quantum rectification might perhaps be present as well.

%\section{Author contributions}

%F.G. conceived and carried out the experiment, analyzed the data, performed the calculations, and wrote the manuscript. M.M. took part in the early stage of measurements, contributed to the cryogenic setup and to writing the manuscript. J.T.P. designed and fabricated the samples, and contributed to writing the manuscript. J.P.P. took part in the early stage of measurements, contributed to the cryogenic setup, took part in the interpretation of the data, and contributed to writing the manuscript. F.G. and J.P.P. discussed the results and implications, and commented on the manuscript at all stages equally.

\end{document}